\documentclass[aps,prx,amsmath,floats,floatfix,twocolumn,
  superscriptaddress,nofootinbib,showpacs]{revtex4-1}

\pdfoutput=1 

\usepackage[colorlinks, pdfborder={0 0 0}, plainpages=false]{hyperref}
\usepackage{graphicx}
\usepackage{xspace}
\usepackage[usenames,dvipsnames]{color}
\usepackage{amssymb}
\usepackage{mathrsfs}
\usepackage{microtype}
\usepackage{etoolbox}

\usepackage[normalem]{ulem}

\newcommand{\eq}[1]{Eq.~\eqref{Eq:#1}}

\newcommand{\Caltech}{\affiliation{TAPIR,
    Walter Burke Institute for Theoretical Physics,
    California Institute of Technology, Pasadena, CA 91125, USA}}
\newcommand{\Cornell}{\affiliation{Cornell Center for Astrophysics and
    Planetary Science, Cornell University, Ithaca, New York 14853, USA}}
\newcommand{\LIGOMIT}{
\affiliation{LIGO Laboratory, Massachusetts Institute of Technology, Cambridge,
Massachusetts 02139, USA}}
\newcommand{\LIGOCIT}{
\affiliation{LIGO Laboratory, California Institute of Technology, Pasadena,
California 91125, USA}}

\begin{document}

\title{Black hole ringdown: the importance of overtones}

\author{Matthew~Giesler}\email{mgiesler@tapir.caltech.edu}\Caltech
\author{Maximiliano~Isi}\thanks{NHFP Einstein Fellow}\LIGOMIT\LIGOCIT
\author{Mark~A.~Scheel}\Caltech
\author{Saul~A.~Teukolsky}\Caltech\Cornell
\date{\today}

\begin{abstract}
  It is possible to infer the mass and spin of the remnant black hole
  from binary black hole mergers by comparing the ringdown gravitational wave
  signal to results from studies
  of perturbed Kerr spacetimes. Typically these studies are
  based on the fundamental quasinormal mode of the
  dominant $\ell=m=2$ harmonic. By modeling the ringdown of accurate
  numerical relativity simulations, we find, in agreement with
  previous findings,
  that the fundamental mode alone
  is insufficient to recover the true underlying mass
  and spin, unless the analysis is started very late in the ringdown.
  Including higher overtones associated with this $\ell=m=2$ harmonic
  resolves this issue, and provides an unbiased estimate
  of the true remnant parameters. Further, including overtones
  allows for the modeling of the ringdown signal for all times
  beyond the peak strain amplitude, indicating that
  the linear quasinormal regime starts much sooner than previously expected.
  This implies that the spacetime is well described as a linearly perturbed
  black hole with a fixed mass and spin as early as the peak.
  A model for the ringdown beginning at the peak strain amplitude can exploit
  the higher signal-to-noise ratio in detectors, reducing uncertainties in
  the extracted remnant quantities.
  These results should be taken into consideration when testing the no-hair theorem.
\end{abstract}


\maketitle

\section{Introduction}\label{sec:intro}
The end state of astrophysical binary black hole (BBH) mergers is a perturbed
single black hole (BH) characterized by two parameters: the final remnant mass
$M_f$ and spin angular momentum $S_f$~\cite{Israel1967,Carter1971,Hawking1972}.
The perturbed BH radiates gravitational waves at a specific set of frequencies
over characteristic timescales completely determined by the mass and spin. The
segment of the gravitational wave signal associated with the single BH's
oscillations is known as the `ringdown' phase, as the perturbed BH rings down
analogous to a struck bell.  The set of frequencies and damping times
associated with a given BH are known as quasinormal modes (QNMs), the damped
oscillations connected to the underlying BH geometry.
The modes can be
decomposed into spin-weighted spheroidal harmonics with angular indices
$(\ell,m)$~\cite{Teukolsky:1972my,Teukolsky:1973,Press:1973}.
For each $(\ell,m)$, there exists a discrete set of complex
frequencies denoted $\omega_{\ell m n}$, where $n$ is the `overtone' index. The
oscillatory behavior is described by $\Re(\omega_{\ell m n})$, while
$\Im(\omega_{\ell m n})$ is related to the damping timescale by $\tau_{\ell m
n} = -\Im(\omega_{\ell m n})^{-1}$.  For a given $(\ell,m)$, the overtone index
sorts the QNMs in order of decreasing damping timescales, so that $n=0$
corresponds to the least-damped mode (i.e. the longest-lived mode), which is
often referred to as the fundamental mode.

The recent detections of merging BBHs~\cite{gw150914,gw151226,o1bbh,gw170104,gw170608,gw170814,gwtc1:2018} by Advanced LIGO \cite{aLIGO} and
Virgo \cite{Virgo},
including the ringdown phase, have stimulated significant interest in measuring
the QNMs from the observations~\cite{gw150914_tgr,Nagar:2016iwa,Cabero:2017avf,
Thrane:2017lqn,Brito:2018rfr,Carullo:2018sfu,Carullo:2019flw}.  Accurately determining the QNMs
allows for precise tests of general relativity
(GR)~\cite{Dreyer:2003bv,Berti:2005ys,Gossan:2011ha,Meidam:2014jpa,Berti:2015itd,Berti:2016lat,Baibhav:2018rfk}.
In~\cite{gw150914_tgr}, the frequency and damping
time of the fundamental mode were inferred from the ringdown data of
the first event (GW150914). The analysis was performed at several
  time offsets with respect to the time of peak strain amplitude.
For sufficiently late values of this start time, the frequency and damping time
were found to be in
agreement with the prediction from GR for a
remnant consistent with the full waveform.
The multiple start times used in
  the analysis reflect an uncertainty
  about when the fundamental mode becomes a valid description for the ringdown,
  as there is noticeable disagreement between the measured mode and the GR
  prediction at early times.
This raises the question:
\textit{at what point in the ringdown does perturbation theory become relevant?}

In this paper, we consider the contribution of QNM overtones to the ringdown.
Including overtones allows for an excellent description of
the waveform well before the fundamental mode becomes dominant
and extends the regime over which perturbation theory is applicable
to times even before the peak strain amplitude of the waveform.
Moreover, an improved model for the ringdown through the
inclusion of overtones can provide more accurate estimates of the remnant
mass and spin~\cite{Buonanno:2006ui,Baibhav:2017jhs}.
Furthermore, the inclusion of higher overtones provides a means
to test GR at a more stringent level, because the
QNM frequencies of all included overtones are independently
constrained by GR for any given $M_f$ and $S_f$.

We begin by demonstrating the benefits of including
overtones, in agreement with~\cite{Buonanno:2006ui, Baibhav:2017jhs},
by analyzing a numerical relativity (NR) waveform. We then show
how overtones can improve the extraction of information from noisy LIGO or Virgo data.
We show that the overtones are not subdominant as is often assumed, but are
instead critically necessary to properly model the linear ringdown regime.
The inclusion of QNM overtones provides a high
accuracy description of the ringdown as early as the time of
peak strain amplitude, where the high signal-to-noise ratio (SNR) can
be exploited to significantly reduce uncertainty in the extracted remnant properties.

\section{Previous studies}\label{sec:prev_work}
There have been numerous attempts to identify the start time of
ringdown, that is, the point in time where a transition has occurred
from the non-linear regime into one where the signal can be described
by a linear superposition of damped sinusoids~\cite{Kamaretsos:2011um,London:2014cma,Thrane:2017lqn,Baibhav:2017jhs,Bhagwat:2017tkm,Carullo:2018sfu}.
To highlight the existing disagreement in the literature, 
the following studies, each using NR waveforms as a testbed,
come to different conclusions regarding this transition time.
In~\cite{Kamaretsos:2011um}, the start of the ringdown phase is inferred
to be $10 M$ (where $M$ is the total binary mass, and $G=c=1$) after
the peak luminosity of the $\ell=m=2$ component
of the strain $h$; this is the time at which the frequency of the
$\ell=m=2$ mode roughly agrees with that of the fundamental QNM.
In~\cite{London:2014cma}, the ringdown portion of the waveform is
considered to be $10 M$ after the peak luminosity of the Newman-Penrose
scalar $\Psi_4$ (related to two time-derivatives of $h$).
A ringdown model with the fundamental and the first two overtones
was built under this assumed start time
and employed in~\cite{Carullo:2018sfu},
which concluded that a start time of $16 M$ after the peak
strain amplitude is optimal.
The peak of $\Psi_4$ is implicitly used as the start time for the
ringdown in~\cite{Baibhav:2017jhs}, where a superposition of the fundamental
mode plus the first two overtones provides an accurate representation of
the remnant properties and the fundamental
frequency expected from perturbation theory.
Interestingly, in one of the earliest analyses of BBH waveforms
using NR simulations, despite the limited numerical accuracy available
for simulations at that time,
Buoananno, Cook, and Pretorius~\cite{Buonanno:2006ui}
were able to fit 3 overtones to the NR ringdown waveform by
extending their analysis to times \textit{before} the peak
amplitude of $\Psi_4$. A superposition of QNMs, including overtones
and pseudo QNMs, became an integral part of modeling the merger-ringdown regime
in earlier EOB models~\cite{Pan:2013rra,Taracchini:2013rva,Babak:2016tgq}.%
\footnote{A recent extension of EOB, referred to as pEOBNR~\cite{Brito:2018rfr},
  was designed for future tests of the no-hair theorem by measuring
  the frequencies of the $\ell=m=2$ and $\ell=m=3$ fundamental modes.
  Restricted to non-spinning binaries, pEOBNR models the
  full inspiral and merger with an attached ringdown model (including overtones),
  in order to avoid deciding at what time the QNMs
  alone provide an accurate description of the waveform.}

A likely cause of confusion is
that start times are defined with respect to the peak
of some waveform quantity, and different authors choose different
waveform quantities for this purpose. To
illuminate the implicit time offsets
incurred by differences in this choice,
consider as a
specific example the GW150914-like NR waveform SXS:BBH:0305
in the Simulating eXtreme Spacetimes (SXS)
catalog~\cite{SXSCatalog,Mroue:2013PRL}. For this waveform,
the peak of $h$ occurs first,
followed by the peak luminosity of $h$, then the peak of $\Psi_4$,
and finally the peak luminosity of $\Psi_4$. These last 3 times are
${}\sim 7 M, 10 M, 11 M$ after the peak of $h$.
As we will show, overtones beyond $n\sim2$ are expected to have
significantly decayed by the peak of $\Psi_4$, so that relying on the
peak of $\Psi_4$ to begin a ringdown analysis may be problematic.

The miscellany of start times above can be reconciled, to
some extent, by considering the contribution of overtones to the
ringdown. Relying solely on the fundamental mode as a description
for the ringdown should result in only late time agreement.
Additional consideration of overtones at late times should result
in finding significantly reduced amplitudes in any overtones
that remain. 
As we demonstrate below, this is because overtones decay more
quickly for larger $n$; each additional included overtone 
leads to a superposition of QNMs that provides a description of the ringdown
at earlier times.
Ignoring the contribution of overtones, by considering
them negligible as in~\cite{Thrane:2017lqn} 
indirectly leads to the conclusion that
remnant properties remain unconstrainable
even in the infinite SNR limit---which we find to be untrue.

\section{Model}\label{sec:model}
We use the fundamental QNM and a varying number of overtones to
determine when the linear QNM solution best describes
the $(\ell,m)$ mode extracted from NR simulations.
Throughout, we focus on
the aforementioned astrophysically relevant NR waveform SXS:BBH:0305
in the SXS catalog, which is modeled after the GW150914 event.
The waveform represents a simulated system with a mass ratio of $1.22$,
  where the larger BH has a dimensionless spin $\vec{\chi}=0.33 \, \hat{z}$ and
  the smaller companion BH has dimensionless spin $\vec{\chi}=-0.44 \, \hat{z}$.
The resulting remnant in this simulation has a final
mass $M_f=0.9520 \, M$ and dimensionless spin $\chi_f=S_f/M_f^2=0.6921$.
We explore at what time the linear QNM description provides
not only an optimal fit for the resulting ringdown waveform,
but also an optimal estimate of the remnant mass and spin.

We model the ringdown radiation as a sum of damped sinusoids~\cite{Vishveshwara1970b,Press1971,Teukolsky,ChandraDetweiler1975} by writing each angular
mode of the complex strain, $h = h_+ - i h_\times$, as
\begin{equation}\label{Eq:qnm}
  h_{\ell m}^{N}(t) = \sum\limits_{n=0}^{N} C_{\ell  mn} e^{-i \omega_{\ell mn} (t - t_0)} \hspace{10pt} t \ge t_0 \;,
\end{equation}
with complex frequencies $\omega_{\ell mn} = \omega_{\ell mn}(M_f,\chi_f)$ as determined by
perturbation theory~\cite{Berti2009,BertiWebsite}.
Here, $t_0$ corresponds to a
specifiable `start time' for the model and times before $t_0$
are not included in the model.
The complex coefficients $C_{\ell mn}$, which are not known a priori as
they depend on the binary configuration and dynamics near merger, are
determined using unweighted linear least squares in the time domain.
The complex-valued amplitudes can be factored into a real-valued amplitude
and phase, $C_{\ell  mn}= |A_{\ell mn}| e^{-i \phi_{\ell  mn}}$, of which we make direct use in Sec.~\ref{sec:ot_m_chi}.

Throughout, we focus on describing the dominant \emph{spherical}
harmonic mode in the NR simulation, the $\ell=m=2$ mode.\footnote{We have verified the presence and early dominance of overtones
  in other resolvable $(\ell,m)$'s in the NR waveform.}
The natural angular basis in perturbation theory is spin-weighted \emph{spheroidal}
harmonics~\cite{Teukolsky:1972my,Teukolsky:1973,Press:1973},
which can be written as an expansion in spin-weighted spherical harmonics~\cite{Press:1973,NewmanPenrose1966,Goldberg1967,thorne80}.
Decomposing the ringdown into spherical harmonics results in mixing of
the spheroidal and spherical bases between the angular functions with the
same $m$, but different $\ell$'s, and this mixing
increases with $\chi_f$~\cite{Press:1973,Berti:2014fga}.
For the SXS:BBH:0305 waveform,
the $\ell=m=2$ spherical harmonic remains a good approximation for
the $\ell=m=2$ spheroidal harmonic. The
amplitudes of the spheroidal and spherical $\ell=m=2$ modes
differ by a maximum of only $0.4\%$, which
occurs roughly $15 M$ after the peak of $h$. This difference is
significantly smaller at the peak.
The mixing is small because higher $(\ell,m)$ harmonics are subdominant
for this waveform, but in a more general case, these higher harmonics may
play a more important role.

\section{Results}\label{sec:results}

\subsection{QNM overtone fits}\label{sec:fits}

\begin{figure}
  \includegraphics[width=1.0\columnwidth]{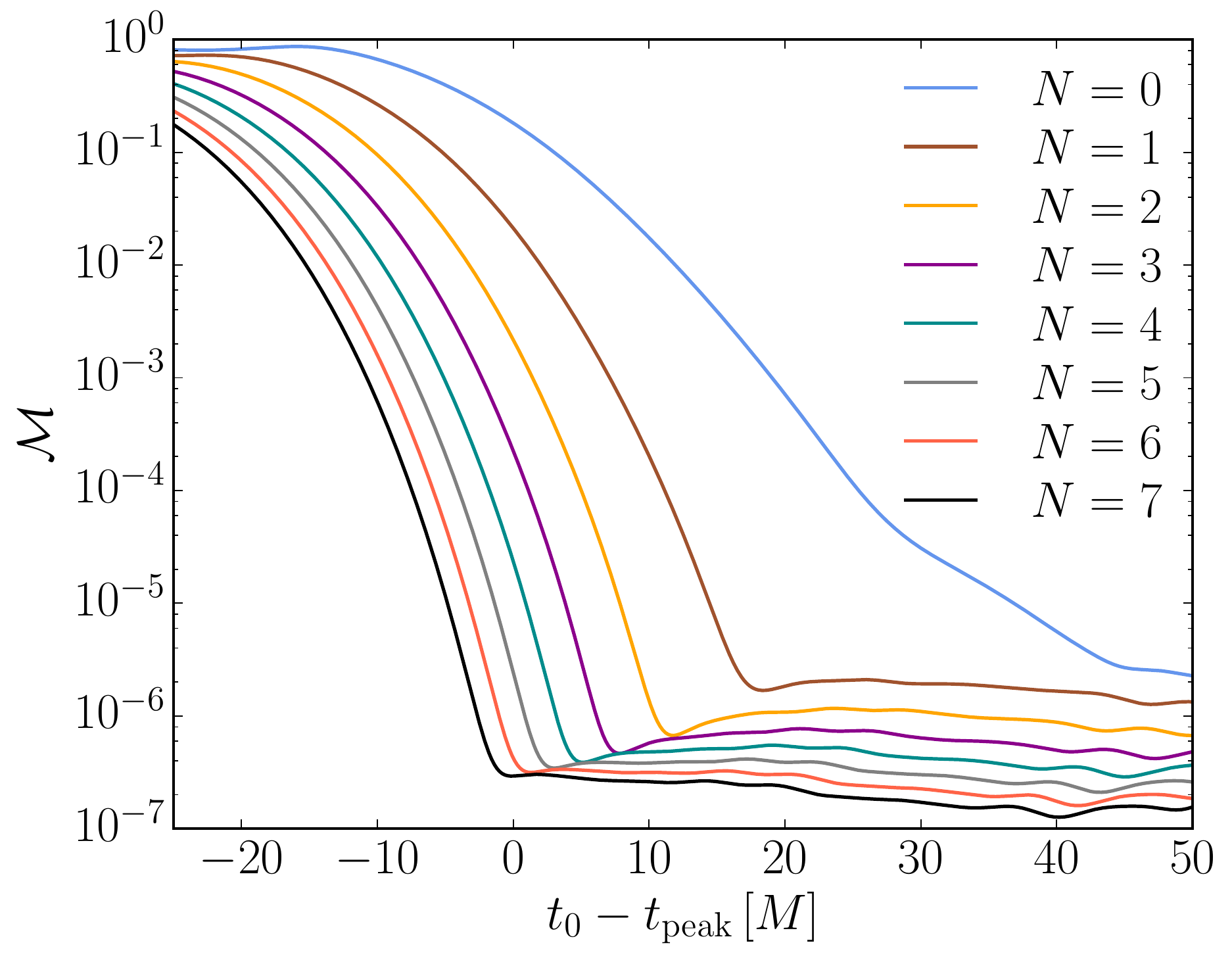}
  \caption{Mismatches as a function of time for the eight models,
    each including up to $N$ QNM overtones. The mismatch associated with
    each model at a given $t_0$ corresponds to the mismatch computed using
    Eq.~\eqref{Eq:mm}, between the model and the NR waveform for $t\ge t_0$,
    where $t_0$ specifies the lower limit used in Eq.~\eqref{Eq:flat_inner_product}.
    Each additional overtone
    decreases the minimum achievable mismatch, with the minimum
    consistently shifting to earlier times.
  }
  \label{fig:mismatch_fot}
\end{figure}

The linear superposition of the fundamental QNM and $N$ overtones is an excellent
description of the waveform around and before the peak strain.
To demonstrate this, we begin by fixing the remnant properties to the final values provided
by the NR simulation. With the mass $M_f$ and dimensionless spin $\chi_f$ fixed,
the set of frequencies $\omega_{22n}(M_f,\chi_f)$ is fully specified
by perturbation theory.
The only remaining free parameters
in Eq.~\eqref{Eq:qnm} are the complex coefficients $C_{22n}$ and
the model start time $t_0$.
For $N$ included overtones, and a given choice of $t_0$,
we determine the ($N+1$) complex $C_{22n}$'s using unweighted linear least squares,
thus obtaining a model waveform given by Eq.~\eqref{Eq:qnm}.
We construct such a model waveform for $t \ge t_0$ at many start times
beginning at $t_0 = t_\mathrm{peak} - 25 M$ and extending to times
$t_0 = t_\mathrm{peak} + 60M$, where $t_\mathrm{peak}$ is the peak
amplitude of the complex strain.
For each start time $t_0$,
we compute the mismatch $\mathcal{M}$ between our model waveform, $h_{22}^{N}$, and 
the NR waveform, $h_{22}^{NR}$, through
\begin{equation}\label{Eq:mm}
  \mathcal{M} = 1 - \frac{\langle h_{22}^\mathrm{NR}, h_{22}^{N} \rangle}
          {\sqrt{\langle h_{22}^\mathrm{NR}, h_{22}^\mathrm{NR} \rangle \langle h_{22}^{N},
              h_{22}^{N} \rangle}} \;.
\end{equation}
In the above, the inner product between two complex waveforms, say $x(t)$ and $y(t)$,
is defined by 
\begin{equation}
  \langle x(t), y(t) \rangle = \int_{t_0}^{T} x(t) \overline{y(t)} \, dt \;,
  \label{Eq:flat_inner_product}
\end{equation}
where the bar denotes the complex conjugate,
the lower limit of the integral is the start time parameter $t_0$ in Eq.~\eqref{Eq:qnm},
and the upper limit of the integral $T$ is chosen to be a time before the NR waveform
has decayed to numerical noise.
For the aforementioned NR simulation, we set $T=t_\mathrm{peak} + 90 M$.

This procedure results in mismatches as a function of $t_0$
for each set of overtones; these are presented in Fig.~\ref{fig:mismatch_fot}.
The figure shows that $N=7$ overtones provides the
minimum mismatch and at the earliest of times, as compared to the other overtone models.
\begin{figure}
  \includegraphics[width=1.0\columnwidth]{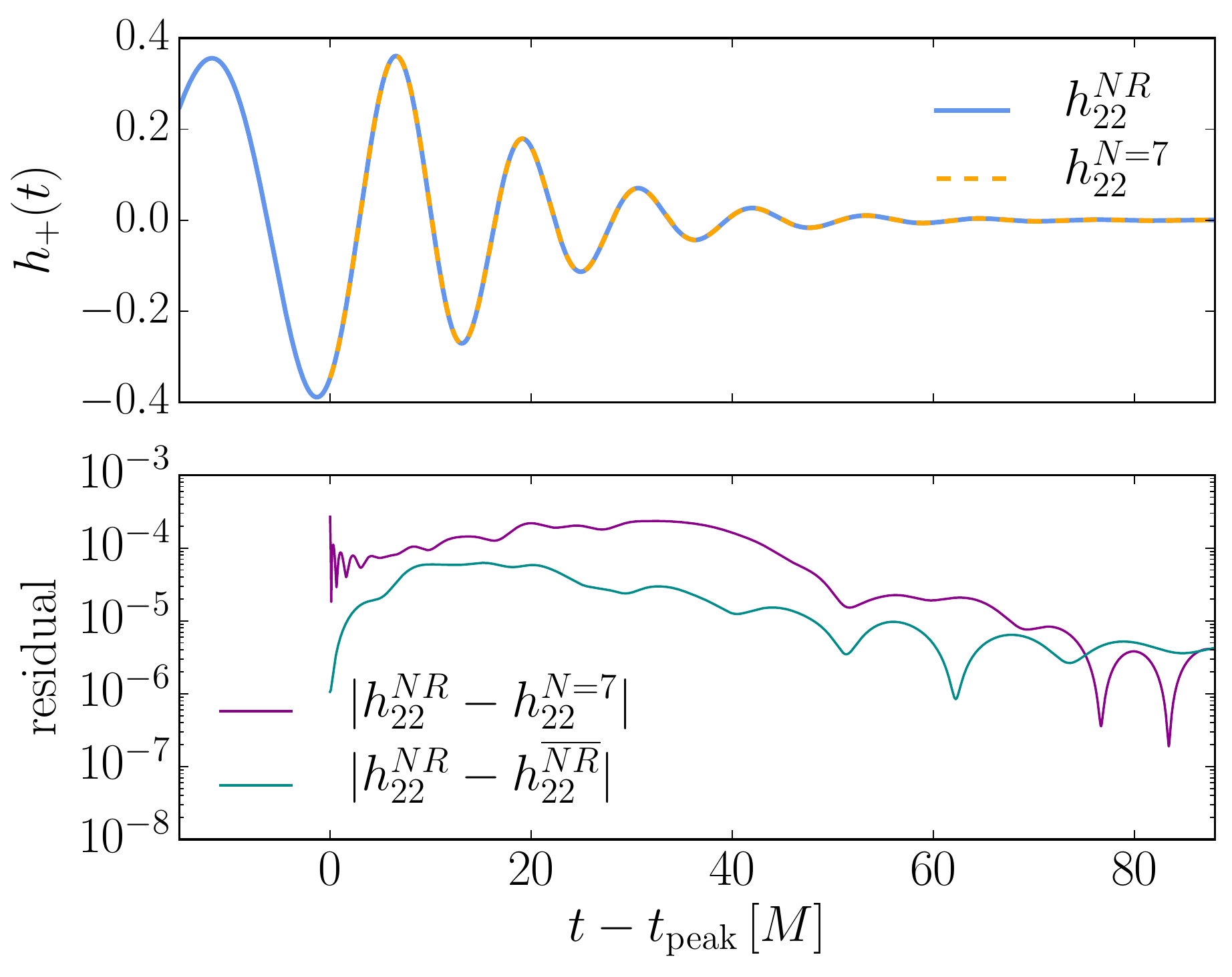}
  \caption{Comparison between the
    plus polarization of the $\ell=m=2$ mode
    of the NR waveform and
    the $N=7$ linear QNM model.
    The QNM model begins at $t_0=t_\mathrm{peak}$.
    The upper panel shows both waveforms, and the
    lower panel shows the residual for $t \ge t_\mathrm{peak}$.
    For reference, the lower panel also shows an estimate of the error in the NR waveform,
      $|h_{22}^{NR} - h_{22}^{\overline{NR}}|$, where $h_{22}^{NR}$ refers to the
      highest resolution waveform of SXS:BBH:0305 and $h_{22}^{\overline{NR}}$ refers
      to the next highest resolution waveform for this same system. The two NR waveforms
      are aligned at $t_0 = t_\mathrm{peak}$, in both time and phase.
  }
  \label{fig:fit_and_resid}
\end{figure}
The waveform corresponding to the $N=7$ overtone model
and $t_0=t_\mathrm{peak}$ is visualized in Fig.~\ref{fig:fit_and_resid},
where the model waveform is compared to the NR waveform along with the fit residual.

At face value, Fig.~\ref{fig:mismatch_fot} provides
us with a guide for determining the
times where a linear ringdown model with $N$ QNM overtones is applicable.
However, relying on the mismatch alone can be deceiving. The $n=7$
overtone decays away very quickly, yet Fig.~\ref{fig:mismatch_fot} shows
that retaining this overtone still produces
small mismatches at times beyond when this mode should no longer
be numerically resolvable.
This is due to overfitting to numerical noise after the higher
overtones in each model have sufficiently decayed.
We find that the turnover subsequent to the first mismatch minimum
in Fig.~\ref{fig:mismatch_fot} is a good approximation for when each overtone
has a negligible amplitude.

\begin{figure}
  \includegraphics[width=1.0\columnwidth]{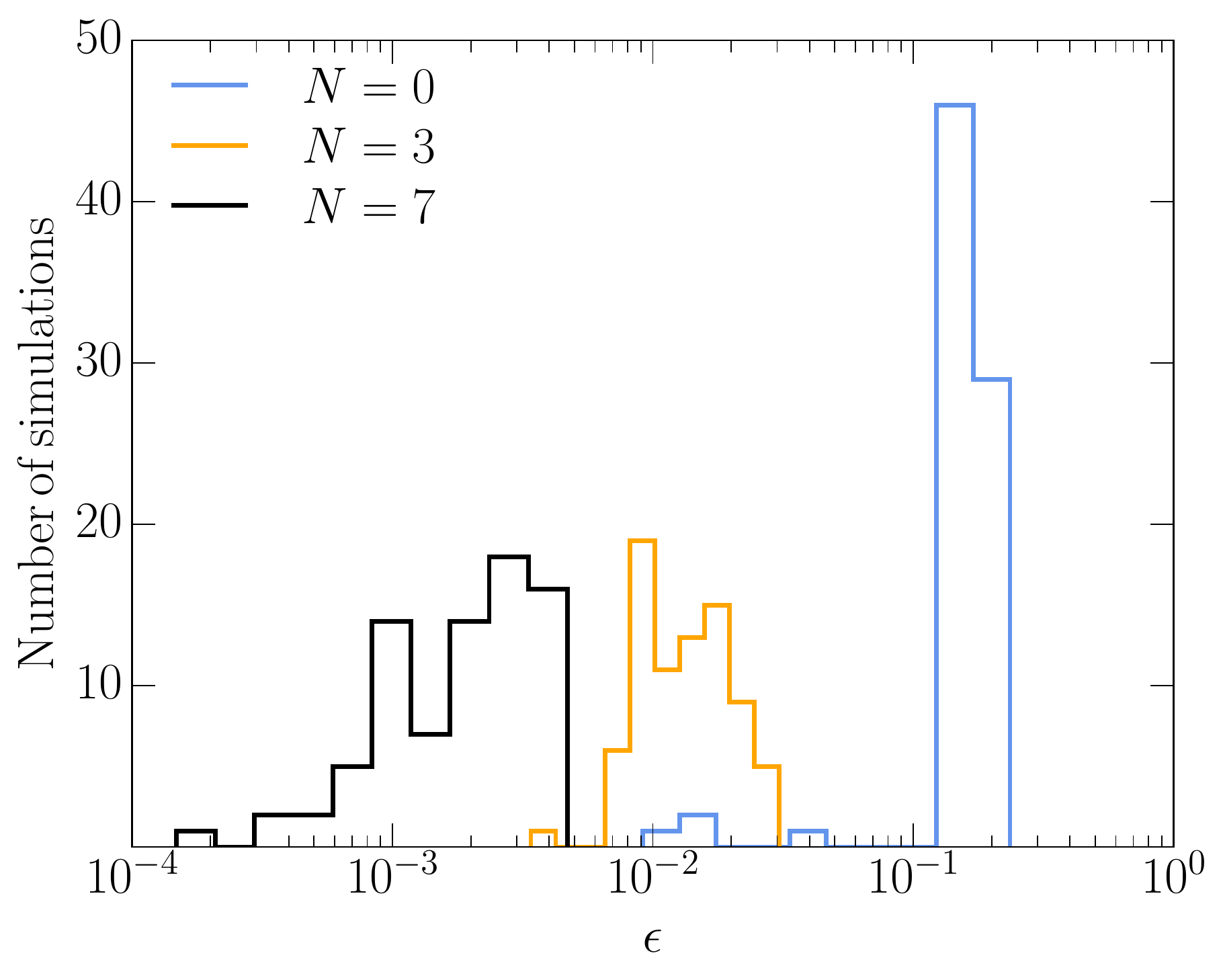}
  \caption{
    The distribution of $\epsilon$, Eq.~\eqref{Eq:epsilon},
    for a range
    of simulations in the SXS catalog. The distribution includes systems with
    mass ratios up to 8 and orbit-aligned component spins with $|\vec{\chi}| \le 0.8$.
    The distributions shown are for $N=\{0,3,7\}$ overtones at the peak of the strain
    amplitude. For the best performing model, $N=7$, the median value
    is $2\times10^{-3}$ and the maximum error in estimating the mass
    and spin is $\sim 5\times10^{-3}$.
    \label{fig:q8_3d}
  }
\end{figure}

It is important then that the model not only minimizes the
residual in the waveform quantity,
but also that it provides faithful estimates of
the underlying system parameters. In particular, we may demand that the
inferred mass and spin agree with the true values known from the NR
simulation.
To check that the model does indeed faithfully represent
the NR waveform with the correct final mass and spin,
we repeat the fits but we allow $M_f$ and $\chi_f$ to vary, and we set
the frequencies of each overtone to their GR-consistent values
through the perturbation-theory formula for $\omega_{22n}(M_f,\chi_f)$.
As a measure of error, we use
\begin{equation}
  \epsilon = \sqrt{(\delta M_f/M)^2 + (\delta \chi_f)^2} \, ,
  \label{Eq:epsilon}
\end{equation}
where $\delta M_f$ and $\delta \chi_f$ are the differences
between the best fit estimates for $M_f$ and $\chi_f$ 
as compared to the remnant values from the NR simulation.
Using a model with $N=7$ overtones and $t_0=t_\mathrm{peak}$,
the best fit estimates for $M_f$ and $\chi_f$
yield a value of $\epsilon$
$\sim 2\times10^{-4}$.
For reference, by comparing the two highest resolutions of this simulation,
  we estimate the error in the NR measured remnant mass and spin
  to be $\delta M_f \sim 1.3\times10^{-5} M$ and $\delta \chi_f \sim 2.1\times10^{-5}$,
  which corresponds to a value of $\epsilon \sim 2\times10^{-5}$.
Furthermore, the difference in the recovered $M_f$ and $\chi_f$ as compared
to the NR values increases as we drop overtones from the model.
This behavior appears to be robust.
Repeating the
above analysis on roughly 80 additional waveforms in the
SXS catalog with aligned spins and mass ratios up to 8~\cite{SXSCatalog,Varma:2018mmi}
yields similar results, with median value of $\epsilon$ $\sim 10^{-3}$.
The full distribution of $\epsilon$ for this part of parameter space,
  with $N=7$ overtones at $t_0=t_\mathrm{peak}$ is shown in Fig.~\ref{fig:q8_3d}.

\begin{figure}
  \includegraphics[width=1.0\columnwidth]{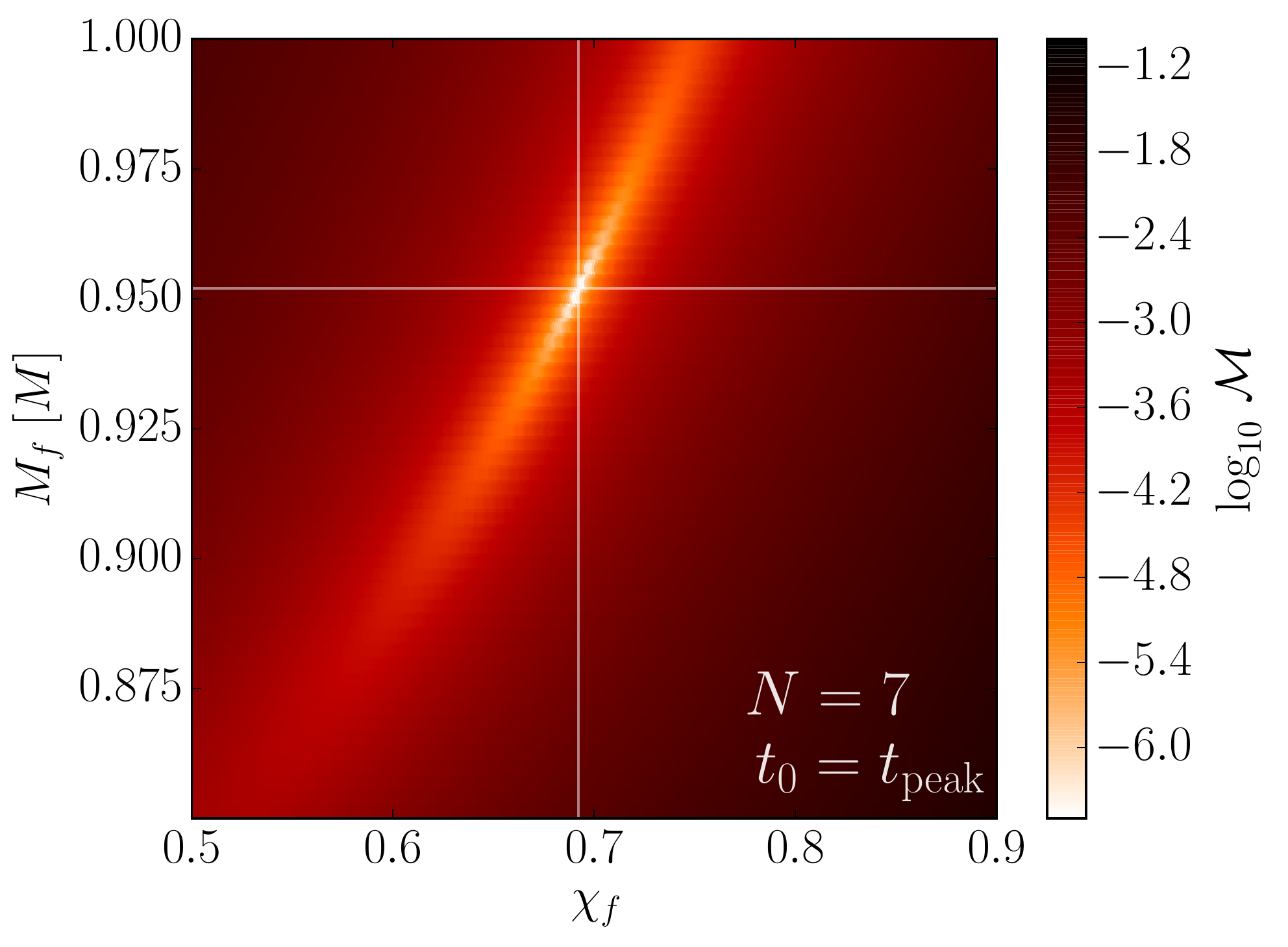}
  \caption{
    Mismatches for the $\ell=m=2$ mode between NR
    and Eq.~\eqref{Eq:qnm} for a grid
    of $M_f$ and $\chi_f$ with $N=7$ and $t_0=t_\mathrm{peak}$.
    The white horizontal and vertical lines correspond to
    the NR values and are in good agreement
    with the $M_f$ and $\chi_f$ mismatch distribution using
    the maximum number of overtones considered.
    \label{fig:N7_tpeak}
  }
\end{figure}

\begin{figure}
  \includegraphics[width=1.0\columnwidth]{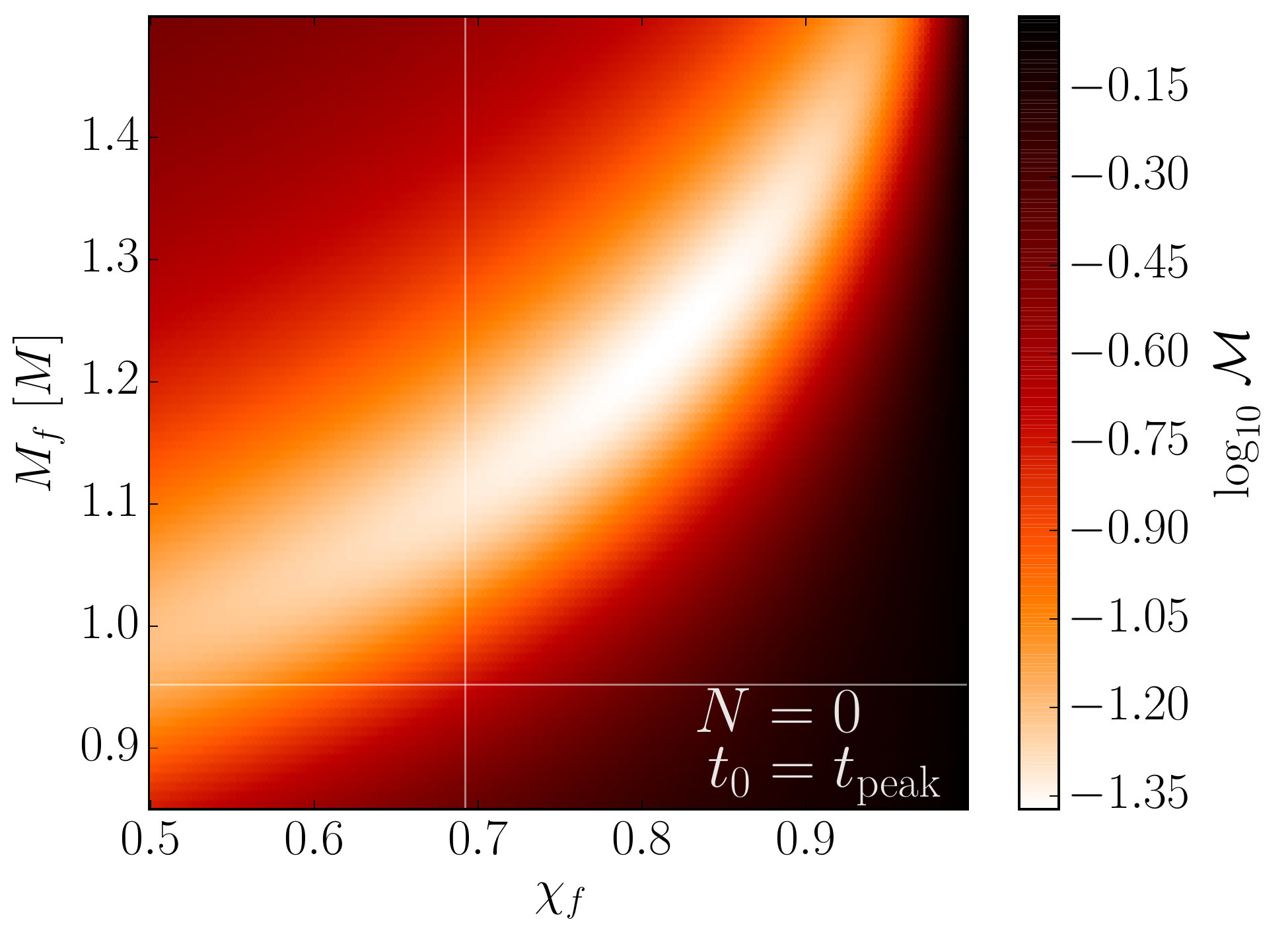}
  \caption{
    The mismatches for the $\ell=m=2$ mode between NR and Eq.~\eqref{Eq:qnm}
    over a grid of $\chi_f$ and $M_f$ with $N=0$, the fundamental
    mode only, and $t_0=t_\mathrm{peak}$.
    The white horizontal and vertical lines correspond to
    the remnant values from NR. As the fundamental mode is subdominant
    at this time, this single-mode model is a poor probe of the underlying
    remnant mass and spin. Note that the mass and mismatch scales used in
    this figure are significantly different than Fig.~\ref{fig:N7_tpeak},
    due to the discrepant single-mode fit at early times.
    \label{fig:N0_tpeak}
  }
\end{figure}

Returning to our analysis of SXS:BBH:0305, to highlight
the worst-fit and best-fit cases and to visualize the mismatch
as a function of mass and spin, we compute the mismatch between NR and
the model Eq.~\eqref{Eq:qnm} with $t_0 = t_\mathrm{peak}$ and
the $C_{22n}$'s determined by a least-squares fit for a grid
of $M_f$ and $\chi_f$ values. 
In Fig.~\ref{fig:N7_tpeak}, we see that with $N=7$ overtones, the mismatch
has a deep minimum associated with the true remnant quantities.
However, using solely the fundamental mode, $N=0$, with
$t_0=t_\mathrm{peak}$ provides largely biased estimates for the remnant
$M_f$ and $\chi_f$, as is visible in Fig.~\ref{fig:N0_tpeak}.
\begin{figure}[h!]
  \includegraphics[width=1.0\columnwidth]{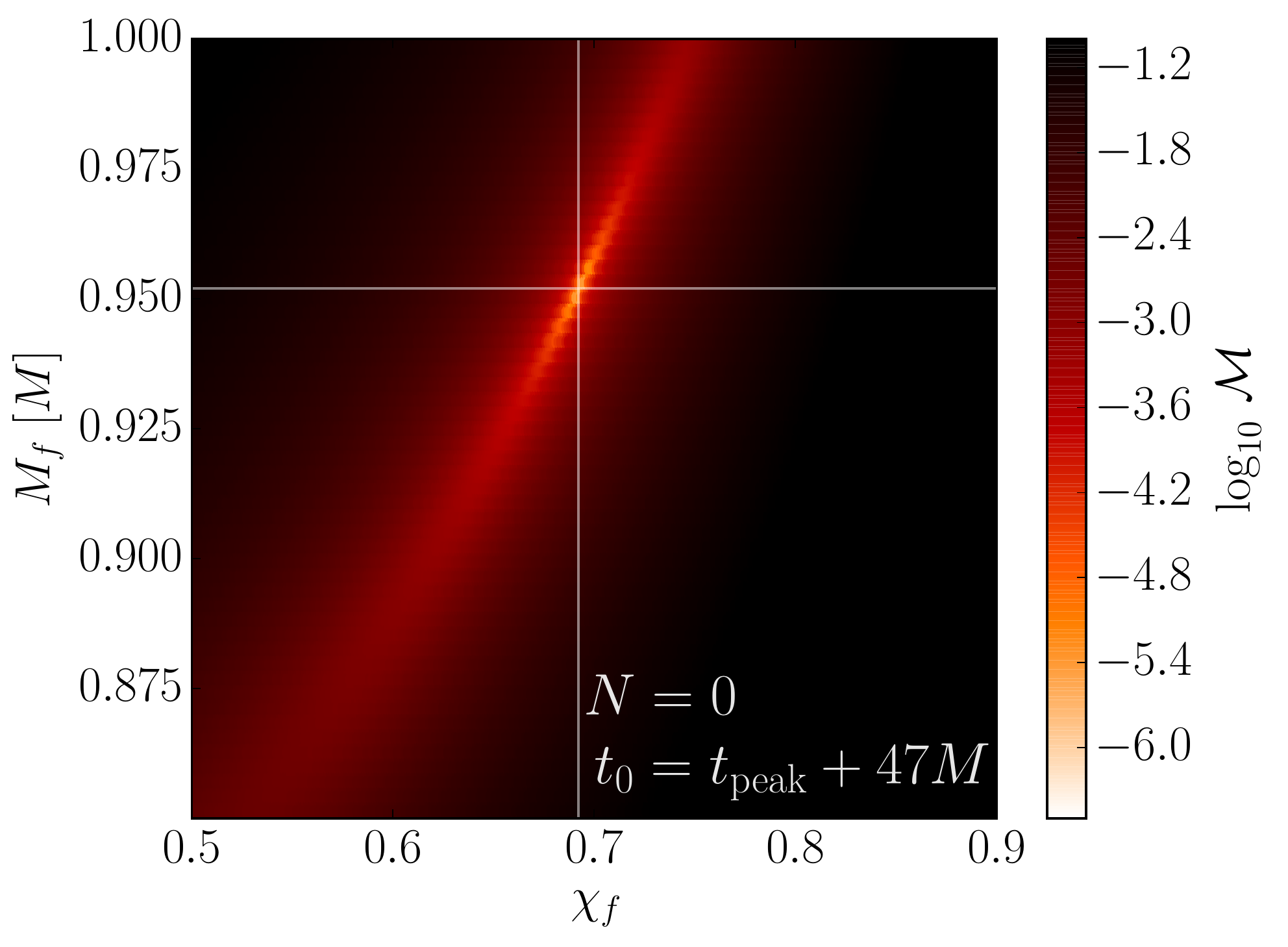}
  \caption{
    Mismatches for the $\ell=m=2$ mode between NR and Eq.~\eqref{Eq:qnm}
    for a grid of $M_f$ and $\chi_f$ with $N=0$,
    the fundamental mode, at a late time $t_0=t_\mathrm{peak} + 47M$.
    Here the fundamental mode is dominant, 
    since the overtones have decayed away by this time.
    Again, the white horizontal and vertical lines correspond to
    the remnant values from NR and now, at this late time, we find good
    agreement between the true values and those recovered by using only
    the fundamental mode as a probe for $M_f$ and $\chi_f$.
    \label{fig:N0_tmin}
  }
\end{figure}
This is not surprising in light of
Fig.~\ref{fig:mismatch_fot}, where at this time the $N=0$ model
provides the poorest mismatch; this is a consequence of the higher
overtones dominating the waveform at this time.
The bias can be overcome by waiting a sufficiently long time,
which allows the overtones to decay away and the
fundamental mode to become dominant.
This can be seen in Fig.~\ref{fig:N0_tmin}, where we repeat
the same procedure with $N=0$ and $t_0=t_\mathrm{peak} + 47M$.
Here the resulting distribution of mismatches in the $M_f-\chi_f$ plane
is on par with the distribution associated with including $N=7$ overtones
and $t_0=t_\mathrm{peak}$, with the $N=7$ case producing a smaller
absolute mismatch than the $N=0$ case. The key point is that we can recover
similar information about the underlying remnant at the peak,
through the inclusion
of overtones, as we can by analyzing the waveform at late times. As discussed in
more detail in Sec.~\ref{sec:ot_m_chi}, 
extending the ringdown model to earlier times allows us to access higher
signal-to-noise ratios and can significantly reduce uncertainties in parameter estimation.

One might be concerned that the additional free parameters in the fit, introduced by including
the overtones, simply allow for
fitting away any non-linearities that may be present, making the fundamental
mode more easily resolvable, and therefore better determining
the underlying
remnant mass and spin. A simple test of this idea is to
repeat the fit while still setting the fundamental
frequencies $\omega_{220}(M_f,\chi_f)$ according to
perturbation theory, but to
intentionally set the frequencies of the overtones to incorrect
values. The fit will then have the same number of degrees of freedom
as previously, but without the correct physics.
Let $\omega_{22n}(M_f,\chi_f)$ be the
set of frequencies determined by perturbation theory and take 
$\widetilde{\omega}_{22n}(M_f,\chi_f)$ to be the set of frequencies with 
the fundamental unmodified, but with
$\widetilde{\omega}_{22n}(M_f,\chi_f) = \omega_{22n}(M_f,\chi_f)(1 + \delta)$, for $n>0$.
As a measure of error, we rely on $\epsilon$, Eq.~\eqref{Eq:epsilon},
  the root-mean-squared error in the estimated mass and spin as compared to
  the known NR values.

\begin{figure}
  \includegraphics[width=0.96\columnwidth]{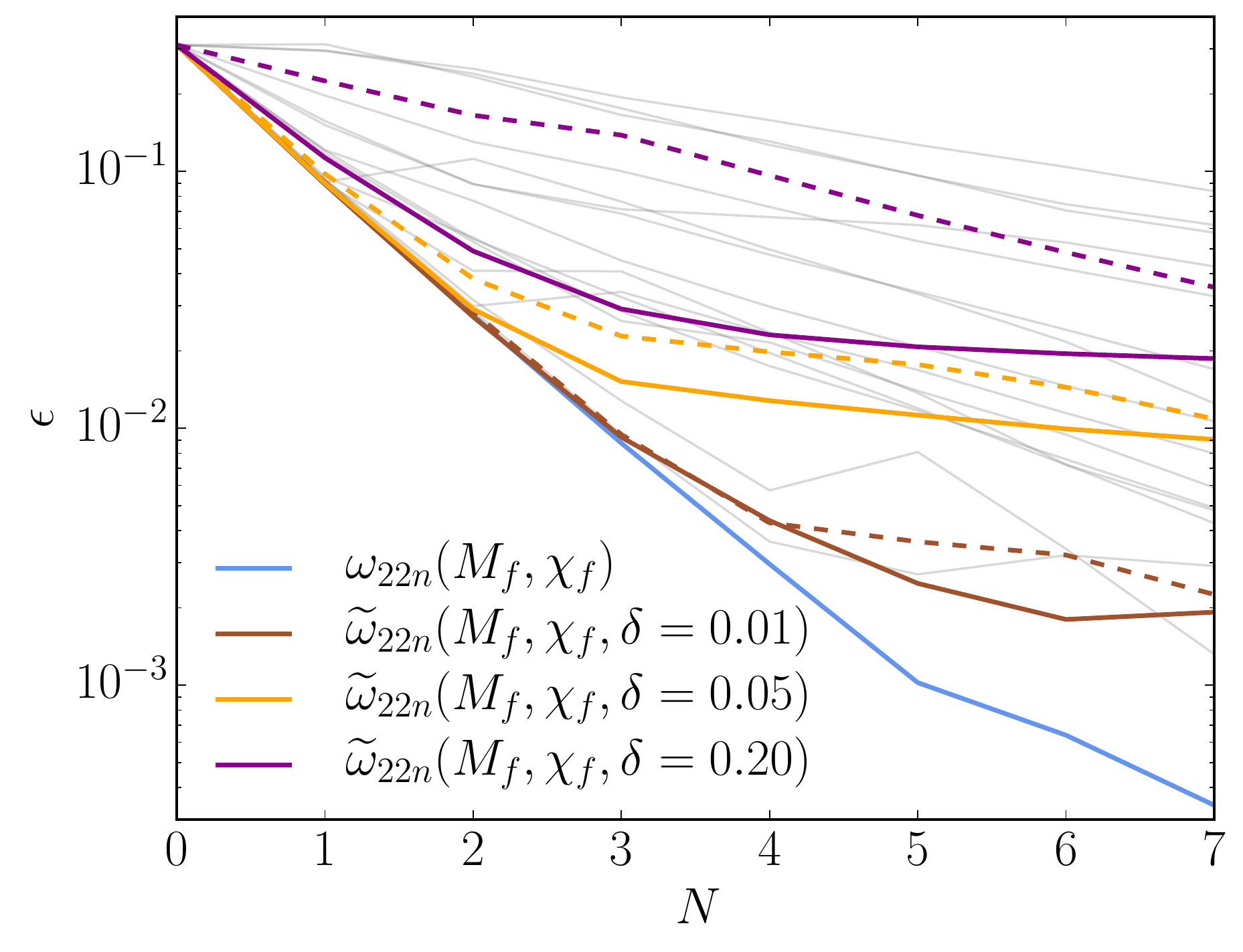}
  \caption{The root-mean-squared error, $\epsilon$, defined in Eq.~\eqref{Eq:epsilon},
    for different sets of frequencies, as a function of number of included overtones $N$. 
    The fits are performed on the spherical $\ell=m=2$
    mode at $t=t_0=t_\mathrm{peak}$.
    The label $\omega_{22n}(M_f,\chi_f)$ represents the set of frequencies consistent
    with perturbation
    theory, while $\widetilde{\omega}_{22n}(M_f,\chi_f)$ represents the set of frequencies
    with the fundamental mode, $n=0$, unmodified but with a slight modification
    to the overtone frequencies by a factor of $(1 + \delta)$.
    For each $\delta$,
      there is an associated dashed line of the same color that corresponds to
      $\delta \rightarrow -\delta$. The faint grey lines correspond to frequencies
      with a random $\delta$ for each $n$, as explained in the last
      paragraph of~Sec.~\ref{sec:fits}.
    The results suggest there is information present in the overtones
    that contribute to extracting the remnant properties at the peak,
    as these outperform a similar set of functions, with the same
    degrees of freedom for each $N$,
    but with frequencies inconsistent with the asymptotic remnant.
    \label{fig:m_chi_err}
  }
\end{figure}

For demonstration purposes, we let
$\delta$ take on values from the set $\pm\{0.01,0.05,0.2\}$ 
and fit to the spherical $\ell=m=2$ mode with
$t_0=t_\mathrm{peak}$ for different numbers of included overtones $N$.
A comparison between the unmodified and modified
models with the same number of degrees of freedom is presented in
Fig.~\ref{fig:m_chi_err}.
From Fig.~\ref{fig:m_chi_err}, it is evident that the unmodified set of
QNMs, $\omega_{22n}(M_f,\chi_f)$, remains true to the underlying
mass and spin and converges to smaller errors as the number of included
overtones is increased. In the case where the overtones are
given slightly incorrect frequencies by the $\delta$ parameter
introduced above, including higher overtones yield fits that remain
biased away from the true values, leading to larger
values of $\epsilon$.

Furthermore, in an additional test we have allowed for
  different values of $\delta$
  for each $n$, each independently sampled from a normal distribution
  with mean $\mu = 0$ and standard deviation $\sigma = 0.2$.
  In this test, each overtone
  frequency is randomly modified to a different extent
  about $\omega_{22n}(M_f,\chi_f)$.
  In all $100$ cases randomly generated from the above distribution,
  the $\epsilon$'s associated with the modified frequencies always
  remain bounded from below by the $\epsilon$ associated with the
  GR frequencies of the asymptotic remnant. A random, representative,
  subset of these $100$ cases is shown as faint grey traces
  in Fig.~\ref{fig:m_chi_err}.
This suggests that the overtones associated with
the asymptotic remnant provide a sufficiently good linear description of
the perturbations for all times beyond the peak of this mode, while
a similar set of overtones that are inconsistent with
the asymptotic remnant do not.

\subsection{Characterizing the overtones}\label{sec:ot_chars}
The behavior in the previous section can be explained by
carefully understanding how the overtones contribute to the
ringdown. As briefly touched on in Sec.~\ref{sec:intro},
the overtones are those modes with $n>0$, where $n$ orders the
modes based on decreasing damping time.
While these modes are the least important in
a time-weighted sense, describing them as `overtones' is somewhat
of a misnomer. In a classical description of harmonics,
overtones are at higher frequencies than the
fundamental, typically multiples of the first harmonic, and
are usually subdominant. However, for QNMs, the overtones
decrease in frequency and are not necessarily subdominant.
As briefly mentioned in Sec.~\ref{sec:model},
the amplitude of each QNM overtone in the ringdown depends on
the binary configuration and the dynamics leading up to merger.
This dynamics specifies the `initial data' for the ringdown, determining
which QNMs are excited and to what extent. As such, the overtone
amplitudes for waveform SXS:BBH:0305 will differ from those
with different `initial data', i.e., binary configurations
with different mass ratios or different spin vectors.

To provide a qualitative understanding of the relative
amplitudes of different overtones, we decompose the ringdown waveform
of SXS:BBH:0305 into its constituent overtones.
Using $t_0=t_\mathrm{peak}$ and $N=7$ overtones, we determine
the $C_{22n}$'s as in Sec.~\ref{sec:fits} with $M_f$ and $\chi_f$ fixed to
the NR simulation values. The corresponding values
$A_n = |C_{22n}(t=t_0=t_\mathrm{peak})|$ form the entries in the
bottom row of Table~\ref{table:amps}. For $N=6$ we keep
$t_0=t_\mathrm{peak}$, so that the amplitudes are measured
with respect to the
peak, but we include in our fit only data for $t \ge t_\mathrm{fit}$,
where $t_\mathrm{fit}$ corresponds to the earliest minimum in
Fig~\ref{fig:mismatch_fot} for this $N$. These amplitudes correspond to
the penultimate row of Table~\ref{table:amps} and the fit time $t_\mathrm{fit}$
is stated, with respect to $t_\mathrm{peak}$, in the last column.
The result of this procedure for the remaining $N$ is Table~\ref{table:amps},
where we provide our best estimate of the amplitudes at $t=t_0=t_\mathrm{peak}$
associated with each overtone.
The values in Table~\ref{table:amps} are computed for the highest
numerical resolution of the NR waveform SXS:BBH:0305, but are truncated
at a level such that the estimates agree with the next highest
resolution.

The initial amplitude 
of the fundamental mode $A_0$ is consistently recovered for all models,
each model having a different $N$ and a different fit time that
is optimal for that $N$. The first few overtones show similar behavior, 
while the higher overtones display larger uncertainties in the recovered
amplitudes and are increasingly sensitive to the fit time
and the number of included overtones.
This sensitivity is a consequence of the strong exponential time dependence in
the overtones and is recognized as the \textit{time-shift problem}~\cite{Dorband:2006gg}.
But, perhaps the most important thing to notice is that the overtones
can have significantly higher amplitudes than the fundamental mode.
As discussed above, the initial amplitudes of the overtones depend on
the details of the nonlinear binary coalescence, which ultimately depend on the
binary parameters.
Consequently, the amplitudes of the overtones relative to the
fundamental mode will vary across parameter space.
The complex amplitudes $C_{\ell m n}$,
also known as the QNM excitation coefficients,
can be written as $C_{\ell m n} = B_{\ell m n} I_{\ell m n}$, where
$B_{\ell m n}$ is a purely geometric piece determined by the remnant
BH, referred to as the QNM excitation factor, and $I_{\ell m n}$ is
the source term that depends
on the binary dynamics ~\cite{Leaver1985,Leaver:1986gd,Berti:2006wq}.
Excitation factors have been computed for the first three overtones
for Kerr BHs in~\cite{Berti:2006wq,Zhang:2013ksa};
these QNM excitation factors can provide some insight into
how the relative amplitudes might behave for different remnant spins.

The NR waveform SXS:BBH:0305 has a dimensionless remnant spin
$\chi_f\sim0.7$, for which the relative excitation
factors, $|B_{22n}|/|B_{220}|$, of the fundamental and the
first three $\ell =m=2$ QNM overtones
are roughly $1.0,3.53,5.23,5.32$.
However, for a remnant of $\chi_f=0$, the excitation factors $|B_{22n}/B_{220}|$ 
of these same QNMs are $1.0,1.28,1.06,0.62$, which indicates
that the overtones may be relatively less important for lower remnant spins.
Using~\cite{Zhang:2013ksa}, we have computed 
the excitation factors for the next two highest overtones
of the remnant of SXS:BBH:0305 and we
find that $|B_{224}|/|B_{220}|\sim15.21$
and $|B_{225}|/|B_{220}|\sim29.31$. Additional excitation factors are
difficult to compute, but the trend is not expected to continue
as it is conjectured that for Kerr BHs $B_{\ell m n} \sim 1/n$
for large $n$~\cite{Berti:2006wq}.

The overtone amplitudes in Table~\ref{table:amps} increase with
overtone number, peak around $n=4$, and then decrease.
Therefore we expect that the rapidly decaying
overtones beyond about $n=7$ are subdominant; this
justifies truncating the expansion in the vicinity of $n=7$.
Prelimary studies indicate that $n=8$ does not improve the
  fit at $t_0=t_\mathrm{peak}$.
An additional caveat is that the amplitudes in
Table~\ref{table:amps} are those recovered from the $\ell=m=2$
spherical harmonic as opposed to the $\ell=m=2$ spheroidal harmonic.
However, the spherical-spheroidal mixing is small
(c.f. Sec.~\ref{sec:model}), and should
not significantly change the qualitative behavior of the
relative amplitudes in Table~\ref{table:amps}.

\begin{table}
  \begin{tabular}{c|l|l|l|l|l|l|l|l|c}
    $N$ & $A_0$ & $A_1$ & $A_2$ & $A_3$ & $A_4$ & $A_5$ & $A_6$ & $A_7$ & $t_\mathrm{fit} - t_\mathrm{peak}$ \\
\hline \hline
0  & 0.971  & - & - & - & - & - & - & - & 47.00 \\
1  & 0.974  & 3.89 & - & - & - & - & - & - & 18.48\\
2  & 0.973  & 4.14 & 8.1  & - & - & - & - & - & 11.85\\
3  & 0.972  & 4.19 & 9.9  & 11.4 & - & - & - & - & 8.05\\
4  & 0.972  & 4.20 & 10.6 & 16.6 & 11.6 & - & - & - & 5.04\\
5  & 0.972  & 4.21 & 11.0 & 19.8 & 21.4 & 10.1 & - & - & 3.01\\
6  & 0.971  & 4.22 & 11.2 & 21.8 & 28 & 21 & 6.6 & - & 1.50 \\
7  & 0.971  & 4.22 & 11.3 & 23.0 & 33 & 29 & 14 & 2.9 & 0.00  \\

  \end{tabular}
  \caption{Best-fit estimates of the
    amplitudes $A_n$ of the fundamental mode
    and overtones in the ringdown of NR simulation SXS:BBH:0305, with $t_0=t_\mathrm{peak}$.
    Amplitudes are computed for
    various values of $N$, the total number of overtones included
    in the fit. Also shown is the time $t_\mathrm{fit}$ where
    the fit is performed
    for each $N$, stated with respect to $t_\mathrm{peak}$.
    $A_n$ are always the amplitudes
    at $t=t_0=t_\mathrm{peak}$, even if the fit is
    performed at a later time.
    The amplitude values
    are truncated such that the last significant figure
    agrees with the the two highest resolutions for
    the NR simulation.
  }
  \label{table:amps}
\end{table}

\begin{figure}
  \includegraphics[width=1.0\columnwidth]{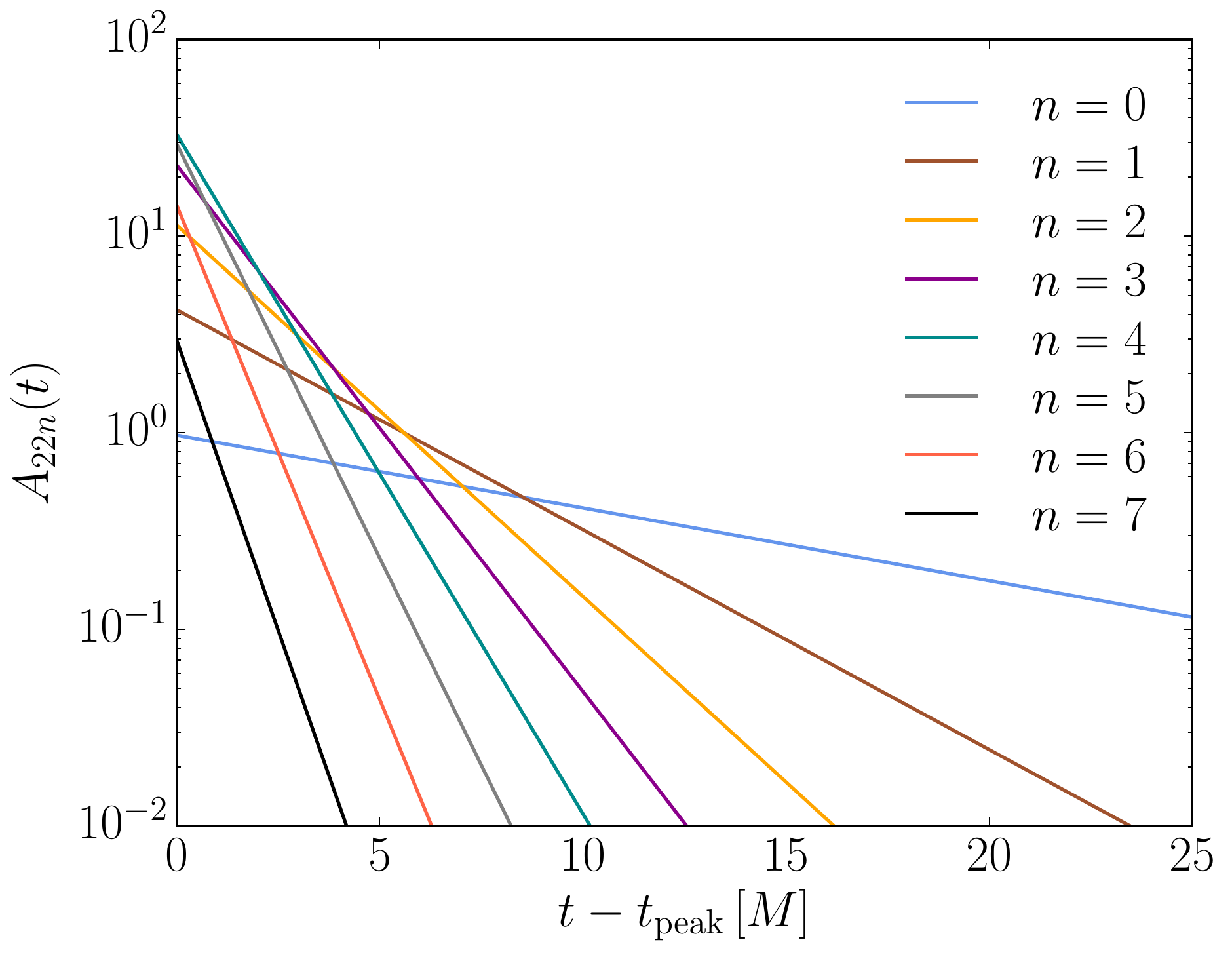}
  \caption{A decomposition of the ringdown in terms of
    the overtones for $N = 7$.
    The evolution of the overtones is computed from
      the analytic decay rates with initial amplitudes
      at $t=t_0=t_\mathrm{peak}$ specified by the bottom row
      of Table~\ref{table:amps}.
    Notice that the fundamental mode does not dominate
      the ringdown of SXS:BBH:0305 until roughly $10 M$
      after $t_\mathrm{peak}$.
  }
  \label{fig:N_decomp}
\end{figure}

Using our results from the last row of Table~\ref{table:amps},
and using the analytic decay rates corresponding to the true $M_f$
and $\chi_f$, we can reconstruct the expected individual contributions
of each overtone to the total $\ell=m=2$ ringdown signal at any given $t$;
in other words, we can compute
the time-dependent amplitudes $A_{22n}(t)$ of each overtone.
These are related to  the $A_n$ in Table~\ref{table:amps} by
$A_{22n}(t) = A_n e^{-(t-t_0)/\tau_{22n}}$.
These amplitudes are shown in Fig.~\ref{fig:N_decomp}.
This establishes why one has to wait until $10-20 M$ after the peak before
the fundamental becomes the dominant contribution.

\begin{figure}
  \includegraphics[width=1.0\columnwidth]{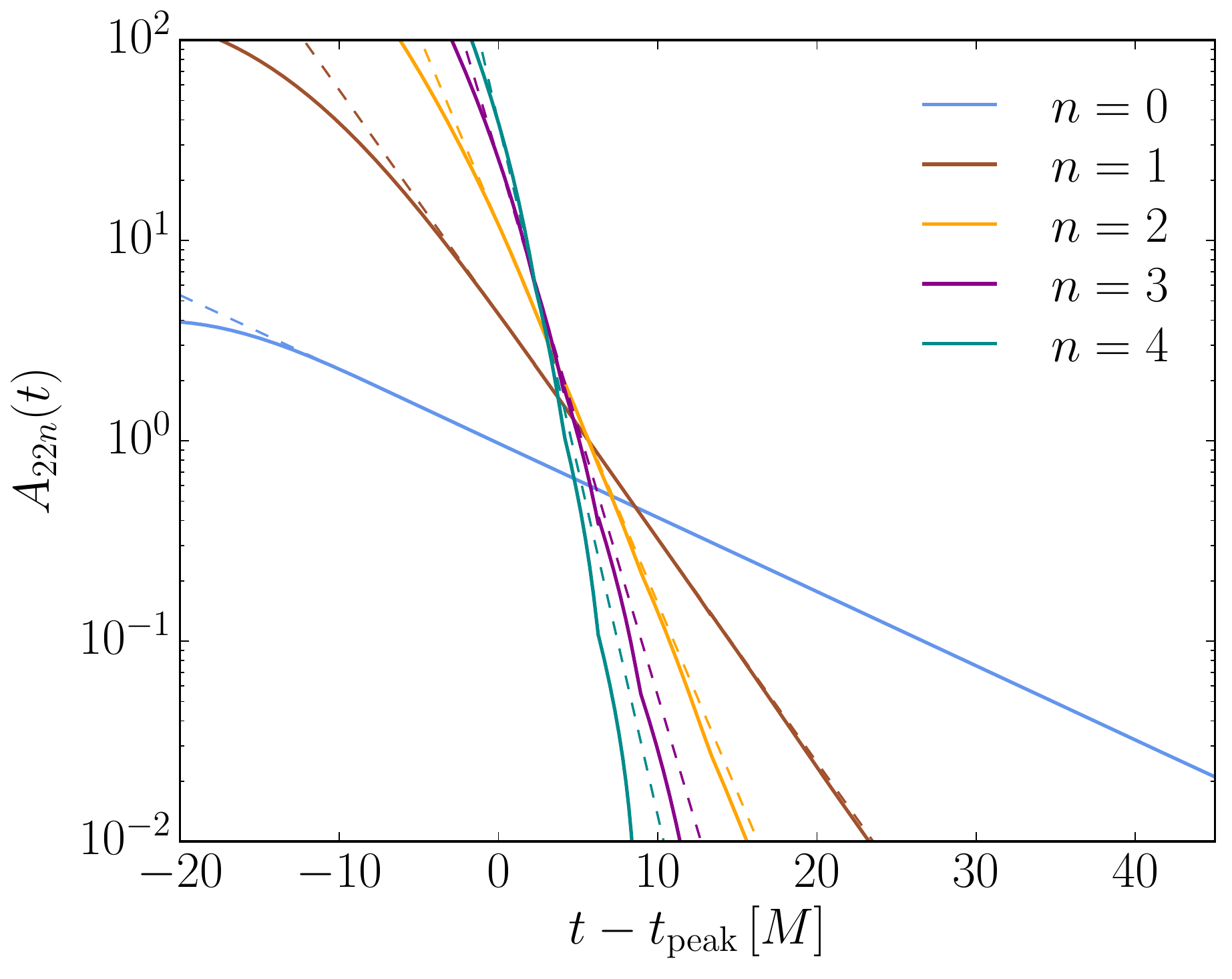}
  \caption{
    The numerically recovered amplitudes for the
    fundamental QNM and the first few overtones at each fit time, $t$
    (solid curves). Dashed lines are the same as the lines
    in Fig.~\ref{fig:N_decomp}.
    The numerically extracted amplitudes across $t$
    agree very well with the expected decay for the longest-lived
    modes, while modes that decay more quickly are more
    susceptible to fitting issues.
    Interestingly, the fundamental mode is in excellent agreement
    with the expected decay rate at times preceding the
    peak amplitude of the strain.
  }
  \label{fig:N_decomp_num}
\end{figure}

Note that Fig.~\ref{fig:N_decomp} uses a single fit
over the range $t\ge t_\mathrm{peak}$,
and assumes the expected analytic time dependence of each overtone
amplitude for $t\ge t_\mathrm{peak}$.  Alternatively, we can attempt to reconstruct each
$A_{22n}(t)$ numerically by performing a different fit for the amplitudes
at each time $t$.
For each time $t$ we choose $t_0=t_\mathrm{fit}=t$ and we fit data only for times $\ge t_0$.
The numerically extracted time dependence of the overtone
  amplitudes, $A_{22n}(t)$, are shown in Fig.~\ref{fig:N_decomp_num}.
Obtaining an accurate fit in this way
is difficult because of various numerical
complications, such as the small differences in frequencies and
amplitudes between neighboring overtones, the poor resolution
of overtones with small amplitudes, and the risk of
overfitting at late times after some overtones have decayed away.
At later times,  there is significantly less
  power in the highest overtones---making them more difficult to resolve.
To mitigate some of these difficulties, when performing the fit
at each time $t$, we exclude overtones whose fitted
amplitude has increased relative to that at the previous time.
This is motivated by the fact that the model is one of
  exponentially damped sinusoids. Therefore, if at any time
  an overtone has a larger amplitude than
  the amplitude recovered at a previous
  time, we consider that overtone
  to no longer be of physical relevance
  and we permanently remove it from the allowed set of modes for
  future fit times.
  It is always the highest overtone available in the remaining
  set of modes that gets dropped, as this mode decays more
  quickly than the other ones.
  Although we only show up to $N=4$ in Fig.~\ref{fig:N_decomp_num}
  because numerically extracting
  amplitudes is difficult at late times,
  the benefit of using overtones up to $N=7$ in estimating the
  remnant mass and spin is apparent in Fig.~\ref{fig:m_chi_err}.
  Consequently, more advanced fitting methods should allow for an
  improvement in numerically recovering higher-order overtones
  as a function of time, which will be explored further in future work.

Finally, it is worth pointing out that there is good agreement between
the model and NR even at times before $t_\mathrm{peak}$, as indicated
by the mismatches in Fig.~\ref{fig:mismatch_fot}, as well as by the
early agreement between the numerically extracted amplitude of the
fundamental mode and the expected analytic behavior visible in
Fig.~\ref{fig:N_decomp_num}. 
Since the QNMs are solutions to perturbed single BH spacetimes,
the agreement could be interpreted as an indication that the region
of the pre-peak waveform already begins to behave as a perturbed
single BH to observers at infinity. This observed behavior will
be explored further in future work.

\subsection{Observing overtones with GW detectors}\label{sec:ot_m_chi}

\newcommand{\rhord}{42}

Overtones can enhance the power of gravitational wave detectors to probe the
ringdown regime. We illustrate this by studying the simulated output of a
LIGO-like detector in response to the same GW considered above, the NR
simulation SXS:BBH:0305. For simplicity, we assume the orbital
plane of the source faces the
instrument head-on (no inclination). We choose a sky location for which the
detector has optimal response to the plus polarization but none to cross, with
polarizations defined in the same frame implicitly assumed in \eq{qnm}.
To mimic GW150914, we rescale the NR template to correspond to a total initial
binary mass of $72 M_\odot$, in the detector frame, and a source distance
of $400 \, \mathrm{Mpc}$. We
inject the $\ell=m=2$ mode of the signal into simulated Gaussian noise corresponding to the
sensitivity of Advanced LIGO in its design configuration
\cite{aLIGO_design_sensi}. This yields a post-peak optimal SNR of ${\sim}\rhord$.%
\footnote{Defined as the SNR in frequencies above 154.68 Hz, the instantaneous frequency at the peak of the time-domain signal.}

\newcommand{\strainpeak}{t_{h\text{-peak}}}

To extract information from the noisy data, we carry out a Bayesian analysis
similar to that in \cite{gw150914_tgr, bayesian_qnm} but based on the overtone
ringdown model of Eq.~\eqref{Eq:qnm}, with $\ell=m=2$ and varying $N$. For any
given start time $t_0$, we obtain a posterior probability density over
the space of remnant mass and spin, as well as the amplitudes and phases of the
set of QNMs included in the template. We parametrize start times via $\Delta t_0 =
t_0 - \strainpeak$, where $\strainpeak$ refers to the signal peak at the detector ($\strainpeak \approx t_\mathrm{peak}-0.48\,\mathrm{ms} \, \approx t_\mathrm{peak}-1.3 \, M$).
Unlike \cite{gw150914_tgr}, we sample over the
amplitudes and phases directly, instead of marginalizing over them
analytically, and we place uniform priors on all parameters. In particular, we
consider masses and orbit-aligned spins within $[10,\,100]\,M_\odot$ and $[0,\,
1]$ respectively. We allow the QNM phases to cover their full range, $[0,\,
2\pi]$, but restrict the amplitudes (measured at $t=\strainpeak$) to
$[0.01,\, 250]\,h_\text{peak}$, where $h_\text{peak}=2\times10^{-21}$ is the
total signal peak. This arbitrary amplitude interval fully supports the
posterior in all cases we consider. We assume all extrinsic parameters, like
sky location and inclination, are perfectly known.
We sample posteriors using the Markov chain Monte Carlo (MCMC) implementations in \texttt{kombine} \cite{kombine} and, for verification, \texttt{emcee} \cite{emcee}.

The highest $N$ we consider in our inference
model is $N=3$, as that is the
most we can hope to resolve given the SNR of our simulation. A guiding principle
for two waveforms to be indistinguishable is $\mathcal{M}
< \mathrm{SNR}^{-2}/2$, in terms of the mismatch $\mathcal{M}$ defined in
Eq.~\eqref{Eq:mm}
but with a noise-weighted inner product~\cite{Flanagan1998,Lindblom2008,McWilliams2010b}.
For the
system at hand, this implies that post-merger templates with mismatches
$\mathcal{M} \lesssim 3 \times 10^{-4}$ are effectively identical. If fitting
from the peak on, Fig.~\ref{fig:mismatch_fot} then implies that differences
between $N\geq3$ templates are unmeasurable. We confirmed this empirically by
checking that $N=4$ does not lead to inference improvements with respect to
$N=3$ and only seems to introduce degenerate parameters.  By the same token,
we have also verified that, at this SNR, our results are largely
unaffected by the presence or absence of the next dominant angular mode $(3,2)$
in the injected NR waveform, as its amplitude is an order of
magnitude weaker than that of the dominant $(2,2)$ mode for the chosen
system. At higher SNRs, additional $(2,2)$ overtones 
and/or angular modes (potentially, with their respective overtones) are necessary to keep the modeling error below the statistical error.

\begin{figure}
  \includegraphics[width=1.0\columnwidth]{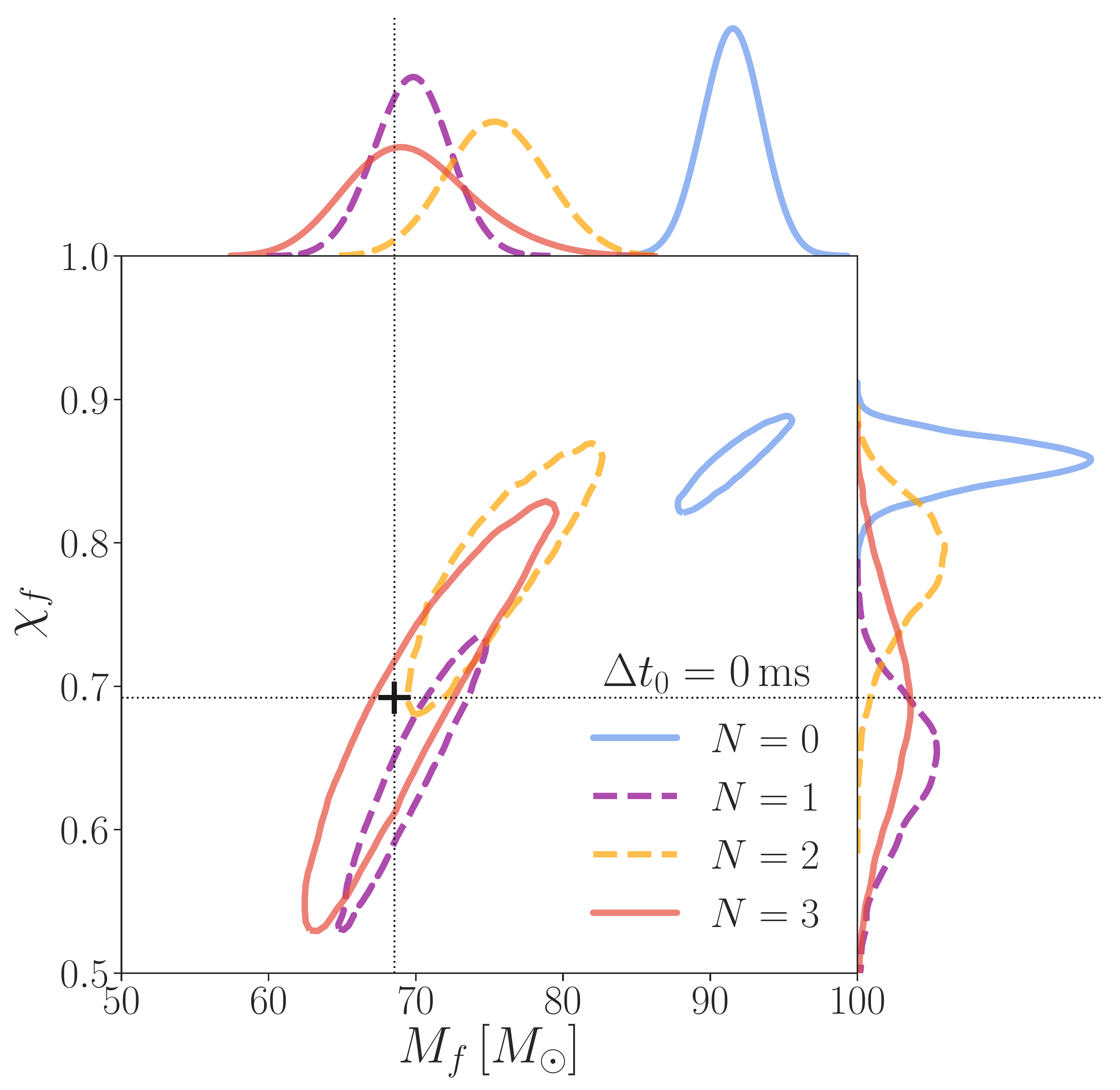}
  \caption{
    Remnant parameters inferred with different number of overtones, using data
    starting at the peak amplitude of the injected strain.  Contours represent
    90\%-credible regions on the remnant mass ($M_f$) and dimensionless spin
    ($\chi_f$), obtained from the Bayesian analysis of a GW150914-like NR
    signal injected into simulated noise for a single Advanced LIGO detector at
    design sensitivity. The inference model was as in Eq.~\eqref{Eq:qnm}, with
    $(\ell=m=2)$ and different number of overtones $N$: 0 (solid blue), 1
    (dashed purple), 2 (dashed yellow), 3 (solid red). In all cases, the
    analysis uses data starting at peak strain ($\Delta t_0 = t_0 -
    \strainpeak = 0$). The top and right panels show 1D posteriors for
    $M_f$ and $\chi_f$ respectively. Amplitudes and phases are marginalized
    over. The intersection of the dotted lines marks the
    true value ($M_f = 68.5 M_\odot$, $\chi_f=0.69$).}
\label{fig:bayes_n} \end{figure}

\begin{figure}
  \includegraphics[width=1.0\columnwidth]{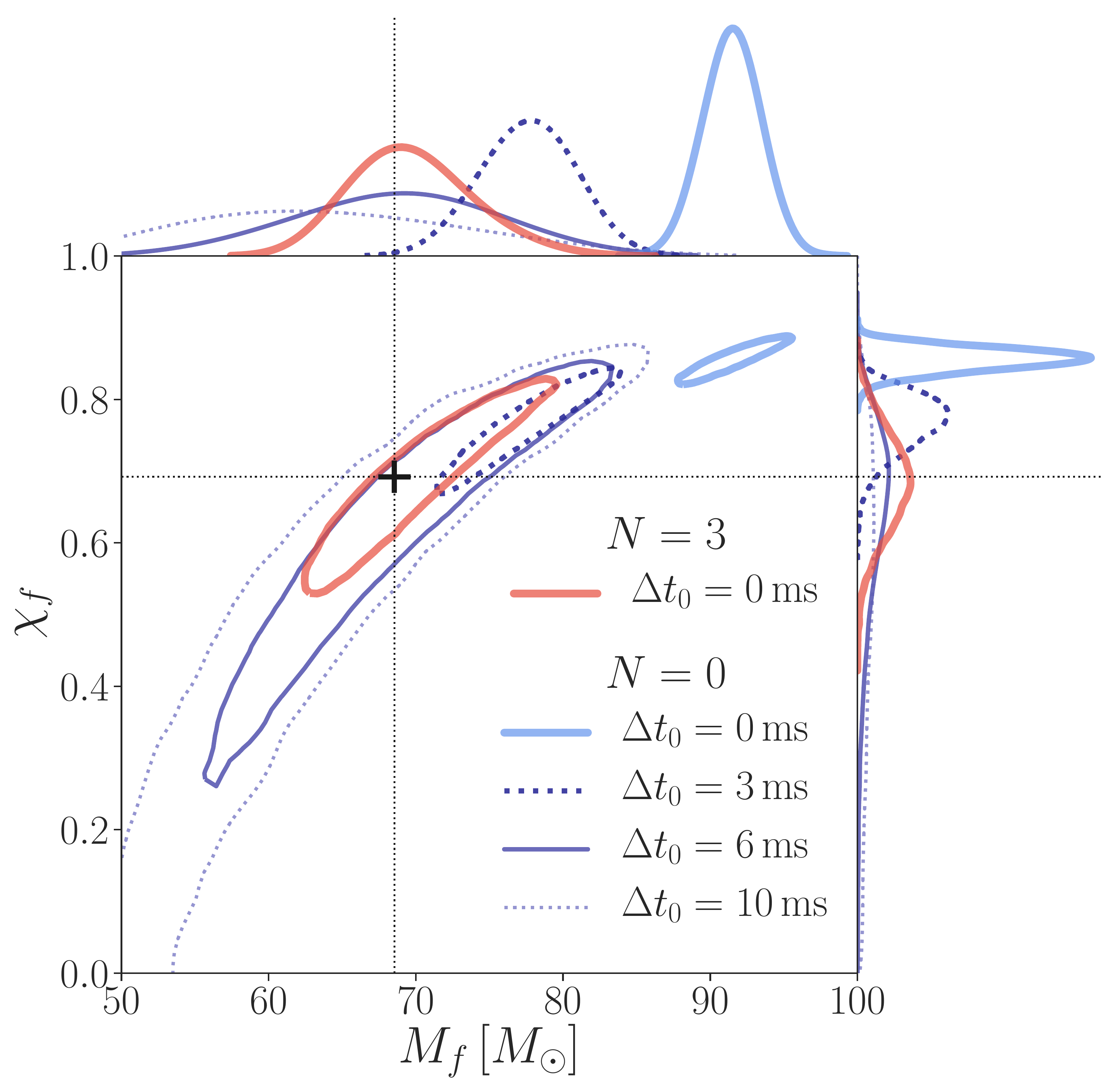}
  \caption{
    Remnant parameters inferred using only the fundamental mode, using data
    starting at different times relative to the peak amplitude of the injected
    strain. Contours represent 90\%-credible regions on the remnant mass
    ($M_f$) and dimensionless spin ($\chi_f$), obtained from the Bayesian
    analysis of a GW150914-like NR signal injected into simulated noise for a
    single Advanced LIGO detector at design sensitivity. For the blue contours,
    the inference model included no overtones ($N=0$) and used data starting at
    different times after the peak: $\Delta t_0 = t_0-\strainpeak \in
    [0,\,3,\,6,\,10]\,\mathrm{ms}$ (blue contours). For the red contour, the
    analysis was conducted with three overtones ($N=3$) starting at the peak
    ($\Delta t_0=0$), as in Fig.~\ref{fig:bayes_n}.  The top and right panels
    show 1D posteriors for $M_f$ and $\chi_f$ respectively. Amplitudes and
    phases are marginalized over. The intersection of the dotted lines marks
    the true value ($M_f = 68.5 M_\odot$, $\chi_f=0.69$). 
  }
  \label{fig:bayes_t0}
\end{figure}

Our findings are summarized in Figs.~\ref{fig:bayes_n} and \ref{fig:bayes_t0}.
In Fig.~\ref{fig:bayes_n} we show the posteriors recovered for the remnant mass
and spin under the assumption that the ringdown begins at the peak of the
signal strain and for models with different numbers of overtones.  For each
case, the main panel displays contours enclosing 90\% of the posterior
probability, while the curves on the top and right represent the corresponding
marginalized distributions for the mass and spin.
As expected, the fundamental mode ($N=0$) is
insufficient to describe the signal near the peak, yielding an estimate of the
remnant properties that is far from the true values determined from the NR
simulation (dotted lines).  As the number of overtones is increased, the
inferred mass and spin become increasingly more accurate, with $N=3$ producing
the best results (true value within top 40\%-credible region).
This result illustrates how the overtones can provide an independent
measurement of the remnant properties by studying the signal near the peak.

We find that the estimate of the mass and spin obtained with overtones at the
peak is more accurate than the one obtained with only the fundamental mode at
later times.  We illustrate this in Fig.~\ref{fig:bayes_t0}, which shows the
90\%-credible regions on $M_f$ and $\chi_f$ inferred using only the fundamental
mode ($N=0$) at different times after the peak strain (blue contours), as well
as the $N=3$ result from Fig.~\ref{fig:bayes_n} for comparison (red contour).
As anticipated in~\cite{gw150914_tgr}, the fundamental mode is a faithful
representation of the signal only at later times, which in our case means that
the true values are enclosed in the 90\%-credible region only for $\Delta t_0
\geq 5$ ms. The penalty for analyzing the signal at later times is a reduction
in SNR that results in increased uncertainty, as evidenced by the large area of
the blue contours in Fig.~\ref{fig:bayes_t0}.  We obtain a more precise
estimate by taking advantage of the overtones at the peak.
We suspect that the observed agreement at $3 \, \mathrm{ms}$ in~\cite{gw150914_tgr}
is a consequence of the lower SNR of GW150914.
At lower SNRs, the statistical errors outweigh the systematic errors
associated with including only the fundamental mode.

\section{Discussion and conclusions}\label{sec:conclusion}
For a given mass $M_f$ and spin $\chi_f$, perturbation theory
precisely predicts the spectrum of QNMs associated with a ringing single BH,
including the characteristic frequencies for these QNMs.
The QNM frequencies are denoted $\omega_{\ell m n}(M_f,\chi_f)$, where $\ell$ and $m$
describe the angular dependence of a mode and $n$, the often-ignored
integer overtone index, sorts QNMs with the same angular dependence by
how quickly they decay. The slowest decaying fundamental mode, $n=0$,
is often considered to be of primary importance, while the more quickly
decaying overtones are often disregarded. However, we find that the overtones
are not necessarily subdominant as is often assumed, but instead, can dominate
the early part of the ringdown.

Using a superposition of QNMs, we
model the ringdown portion of the $\ell=m=2$ mode of the
numerical relativity waveform SXS:BBH:0305,
which is consistent with GW150914.
We find that with enough included overtones,
the QNMs provide an excellent description for the GW strain for all
times beyond the peak amplitude of the complex strain $h$.
For the GW150914-like NR waveform we analyzed,
the overtones dominate the early part of the perturbations but
decay away much more quickly than the fundamental mode, which eventually
becomes dominant roughly $10 M$ after the peak amplitude
(Fig.~\ref{fig:N_decomp}).  This later time where the fundamental dominates
is sometimes
referred to in the literature as
the start of the ringdown,
the time of a transition to the linear regime, or the beginning of the
domain of applicability of perturbation theory. However, this time is merely
the time at which one may ignore the contribution of overtones,
which play a key role in the early ringdown.
Including the QNM overtones extends the reach of
perturbation theory back to the time of the peak strain amplitude,
indicating that the linear ringdown regime
begins much earlier than
one would conclude by ignoring these additional modes.
As mentioned in Sec.~\ref{sec:fits}, we have verified,
on a sizeable set of aligned-spin waveforms in the SXS catalog, 
that the inclusion of overtones provides an accurate
model for the post-peak strain. Not only do the overtones provide
excellent mismatches, but the best fit mass and spin are accurately
recovered with median absolute errors in $M_f/M$ and $\chi_f$ of $\sim10^{-3}$.
We therefore expect the early dominance of overtones to be a generic
feature of the ringdown.

The QNM overtones can enhance the power of GW detectors to probe the ringdown
regime. They can be used to extract information about QNMs at the peak of the
signal, where the SNR is high.  In contrast, the usual approach relies solely
on the later portion of the signal that is dominated by the (initially weaker)
fundamental mode, paying the price of larger statistical errors and
uncertainty in the appropriate time where this mode
dominates~\cite{gw150914_tgr,Nagar:2016iwa,Cabero:2017avf,Thrane:2017lqn,Brito:2018rfr,Carullo:2018sfu,Gossan:2011ha,Meidam:2014jpa,Berti:2015itd,Berti:2016lat,Baibhav:2018rfk,Carullo:2019flw}.
This effect is visible in Fig.~\ref{fig:bayes_t0}, where a model with $N=3$
overtones remains faithful to the true remnant mass and spin with less uncertainty
than one with $N=0$ at later times. The resolvability of
these overtones provides a set of independent modes, each with unique frequencies, that
can potentially be used to constrain deviations from GR.

Studies of the ringdown GW spectrum can provide a direct way to experimentally
determine whether compact binary coalescences result in the Kerr BHs predicted
by GR \cite{Dreyer:2003bv,Berti:2005ys}. This includes tests
of the no-hair theorem and the area law, as well as searches for BH mimickers.
The program, sometimes known as ``black-hole spectroscopy,'' generally requires
independent measurement of at least two modes, which are conventionally taken
to be the fundamentals of two different angular harmonics (e.g.~\cite{Gossan:2011ha,Baibhav:2018rfk}). However, such choice is only
available for systems that present a sufficiently strong secondary angular
mode, which only tends to occur under some specific conditions (e.g. for high mass ratios)
~\cite{Varma:2016dnf,Capano:2013raa,Littenberg:2012uj,Bustillo:2016gid,Varma:2014}.
Further, as we have observed, these fundamental modes
should dominate only at late times, being subject to significantly more noise
than modes than can be extracted near the peak of the waveform.
The extraction of an overtone,
in addition to the fundamental mode, could potentially serve as an alternative
two-mode test of the no-hair theorem.

The impact of overtones on ringdown tests of GR can already be glimpsed from
Fig.~\ref{fig:bayes_t0}: by studying the QNMs at early and late times we may
obtain two independent measurements of the remnant parameters, enabling
powerful consistency checks.  Unlike tests that rely on a multiplicity of angular
modes, studies of overtones should be feasible at SNRs achievable with existing
detectors, as we demonstrate by our study of a GW150914-like signal seen
at design sensitivity by Advanced LIGO (Sec.~\ref{sec:ot_m_chi}).
For signals in which they are measurable, higher
angular modes and their overtones could make these tests even more powerful.
Overtones can therefore enable a whole new set of precision studies of the ringdown and
make black-hole spectroscopy realizable with current detectors.

\begin{acknowledgments}
The authors thank Vijay Varma for many valuable discussions.
We also thank Katerina Chatziioannou and Leo Stein for useful comments.
M.G. and M.S. are supported by the Sherman Fairchild Foundation and NSF
grants PHY-1708212 and PHY-1708213 at Caltech.
M.I.\ is a member of the LIGO Laboratory.
LIGO was constructed by the California Institute of Technology and
Massachusetts Institute of Technology with funding from the National
Science Foundation and operates under cooperative agreement PHY-0757058.
M.I.\ is supported by NASA through the NASA Hubble Fellowship
grant No.\ HST-HF2-51410.001-A awarded by the Space Telescope
Science Institute, which is operated by the Association of Universities
for Research in Astronomy, Inc., for NASA, under contract NAS5-26555.
S.T. is supported in
  part by the Sherman
  Fairchild Foundation and by NSF Grants PHY-1606654 and
 ACI-1713678 at Cornell.
Computations were performed on the
Wheeler cluster at Caltech, which is supported by the Sherman Fairchild
Foundation and by Caltech.
Computation were also performed on the Nemo computing cluster at the University 
of Wisconsin-Milwaukee, supported by NSF Grant PHY-1626190.

\end{acknowledgments}

\bibliography{References,ligo}

\end{document}